\UseRawInputEncoding
\documentclass[12pt]{article}

\usepackage{graphicx}   
\usepackage{hyperref}   

\usepackage[fleqn]{amsmath}
\usepackage{amsmath}
\usepackage{amsfonts}   
\usepackage{amssymb}    
\usepackage{bm}		
\usepackage{enumitem}		
\usepackage{enumitem}
\usepackage{braket}
\usepackage{ulem}

\begin{document} 

\title{Experimental tests of alternate postulates to describe excited-state decay in DeWitt-like and Everett-like versions of quantum mechanics} 
\author{Jon Geist}
\date{2019 Oct 01}

\maketitle

\section*{Abstract}

Simulations of the observed transition rate (decay rate) versus time for excited-state decay in Everett's relative-state formulation of quantum mechanics and DeWitt-like extension of Everett's formulation are described.  It is shown that Everett's formalism is not general enough to describe excited-state decay, and that tritium lifetime measurements rule out all plausible DeWitt-like extensions.  Furthermore, comparison of lifetime measurements and first-principles calculations of excited electronic states of simple atoms limit Everett-like extensions to a small effect for these atoms.  

\section*{Introduction}

In 1957, Everett \cite{Everett01} published a paper in Reviews of Modern Physics, which was based on his much later published PhD thesis \cite{Everett02}. These works described what he called a \lq \lq relative state" formulation of quantum mechanics.  In these works he claimed that all terms in a quantum mechanical superposition existed simultaneously on different non-interacting branches of reality.  He further claimed that observers present on one branch would be unaware of the existence of the other branches. Finally, he concluded that an observation of a quantity described by a superposition of states would give the appearance of a discontinuous change of the wavefunction from a superposition to a single eigenstate on each branch of reality, thereby replacing the so-called \lq\lq collapse of the wave packet" by the splitting of a single branch of reality into multiple, equally-real, temporally-parallel branches of reality.  

Reality branching was the only truly new result of Everett's reformulation of quantum mechanics (EQM).  Furthermore, without considering excited-state decay (spontaneous transitions) in detail, he claimed that 
\begin{quote}
\textit{It is found that experiences of the observer (magnetic tape memory, counter system, etc.) are in full accord with predictions of the conventional "external observer" formulation of quantum mechanics, based on Process 1}. 
\end{quote} Note that Everett defined Process 1 as the collapse of the wave packet following a measurement, which he eliminated from his formalism by proposing \lq \lq to regard pure wave mechanics (Process 2 only) as a complete theory", where Process 2 postulates that 
\begin{quote}
\textit{a wave function that obeys a linear wave equation everywhere and at all times supplies a complete mathematical model for every isolated physical system without exception. It further postulates that every system that is subject to external observation can be regarded as part of a larger isolated system.}
\end{quote}
Also, note that beyond some formal aspects of the formulation, Everett \lq\lq  never clearly spelled out how his theory was supposed to work."\cite{Schlosshauer}.   

To make a long story short, Everett's paper was not well received by the physics community.\cite{Byrne01} Interest in it soon waned until about 1970, when DeWitt\cite{DeWitt1, DeWitt2} claimed, without spelling out how EQM actually worked, that EQM resolved Schroedinger's cat paradox (as translated into English in \cite{Trimmer}). DeWitt's application of EQM to the cat paradox explicitly assumed that reality branching applied to excited-state decay, and this idea seems almost universal among advocates of EQM.    

DeWitt's article \cite{DeWitt1} included a figure whose caption stated, 
\begin{quote}\textit{\textbf{Schroedinger's cat} The animal (is) trapped in a room together with a Geiger counter and a hammer, which upon discharge of the counter, smashes a flask of prussic acid.  The counter contains a trace of radioactive material - just enough that in one hour there is a 50\% chance one of the nuclei will decay and therefore an equal chance that the cat will be poisoned.  At the end of the hour the total wavefunction for the system will have a form in which the living cat and the dead cat are mixed in equal portions. Schroedinger felt that the wave mechanics that led to this paradox presented an unacceptable description of reality. However, Everett, Wheeler, and Graham's interpretation of quantum mechanics pictures the cats as inhabiting two simultaneous, noninteracting, but equally real worlds. }\end{quote}

The figure shows the superposition of a live cat and a dead cat, an intact flask and a broken flask, and a hammer suspended above the flask and in a position following the breaking of the flask to illustrate the simultaneous existence of two non-interacting, but equally real worlds. In one of these worlds, pictured in red, the cat is dead and the broken flask has recorded the decay of one of the nuclei.  In the other world, pictured in black, the cat is alive and the intact flask shows that none of the nuclei have decayed so far.  

Clearly the presence of the cat is not crucial to the assumed branching of one world into two.  Furthermore, the hammer and flask could be replaced by any decay detection and recoding device because Everett stated in \cite{Everett01} 
\begin{quote}\textit{As models for observers we can, if we wish, consider automatically functioning machines, possessing sensory apparatus and coupled to recording devices capable of registering past sensory data and machine configurations.} \end{quote}
Only the radioactive material and a means to detect and record the first decay event is required. For instance, the Geiger counter and a computer monitoring its output would suffice. There is no need for a human to observe the computer output in real time or anytime.  According to Everett, detection and recording are all that is necessary.  

Besides the implicit application without details of Everett's formulation to excited-state decay, another important result of DeWitt's article was a rebirth of interest in Everett's ideas. In fact, it was DeWitt who coined the name \lq\lq many worlds", which is more charismatic than \lq\lq relative state formalism" and which emphasizes the most striking difference between EQM and CQM.  Indeed, DeWitt explicitly stated \cite{DeWitt1}
\begin{quote}\textit{Moreover, every quantum transition taking place on every star, in every galaxy, in every remote corner of the universe is splitting our local world on earth into myriad of copies of itself.} \end{quote}

DeWitt also stated without justification in \cite{DeWitt1}, 
\begin{quote}\textit{Clearly the EWG view of quantum mechanics leads to experimental predictions identical with those of the Copenhagen view.  This, of course, is its major weakness. Like the original Bohm theory $^6$ it can never receive operational support in the laboratory. No experiment can reveal the existence of the \lq\lq other worlds" in a superposition like that in equations 5 and 6.} \end{quote}  
This statement apparently bolstered belief in the superficially plausible idea that a search for laboratory tests of EQM would be a waste of time. Currently, reality branching, including that in DeWitt's extension of EQM to the cat paradox, continues to be treated by many physicists as if it is an untestable but perhaps inevitable consequence of quantum mechanics.\cite{Kent}  

In essence, Everett introduced an entire new branch of physics, quantum \lq\lq kladodynamics" after the Greek word for \lq\lq branch" and then said that it was not subject to study because all of its predictions were in full accord with the predictions of conventional quantum mechanics. Even if this were true (which it isn't as demonstrated in excruciating detail in this report) some properties of quantum kladodynamics can be still be deduced and they violently defy the conservation laws that form the basis of the Standard Model. 

If a single experimental setup in its initial state splits into multiple setups in different final states on multiple branches then the baryon number and the mass/energy (and any other non-zero conserved property on the original branch) are not conserved during the split. Therefore every single branch must be an infinite source of entities and properties that are constrained by conservation laws and uncertainty relations between branching events. When this idea is pursued through Everett's description of multiple observers, it becomes clear that every new branch must exist in an entirely new universe as emphasized by DeWitt.  

It took the universe we know billions of years just to evolve planets, yet quantum kladodynamics spews out vast numbers of almost perfect copies of current versions of each universe instantaneously (or they grow and blossom at the speed of light) with every quantum transition.  Such claims should not be accepted based on some formal manipulations without very serious study.       

Indeed, a closer look at EQM and DeWitt's extension of it to excited-sate decay shows that data to test Everett's formulation and DeWitt-like extensions are already available. The descriptions of the theory, while incomplete, are complete enough to simulate excited-state decay without detailed knowledge of how every aspect of EQM is supposed to work.  Instead, all that is needed are alternate versions of the equations that describe differences between measurement results in branching and non-branching realities.  In other words, we do not need what Tegmark\cite{Tegmark} calls the outside view of our universe, but only the inside view of those who live in it, as discussed in detail later in this report. 

I previously showed \cite{Geist01} that measurable differences are possible for excited-state decay. That work tested whether multiple decays are possible on a single reality branches for which no prior decays were detected.  The answer was yes, if and only if, the relation between the expectation value of the transition rate $\Gamma_A$ measured by any observer and the actual transition rate on any branch $\Gamma_B$ is given by $\Gamma_A = (1- \epsilon) \Gamma_B$, where $\epsilon$ is a real number between 0 and 1 that is independent of time. (I know of no other study rigorously applying EQM to excited-state decay.) 

This result is a lot more important than I appreciated at the time.  It is the first prediction of a version of quantum kladodynamics that can be tested. Iodine-131 ($^{131}$I) with a half life of 8 days and a stable xenon decay product would make an excellent test case.  If multiple decay events are forbidden on any single history (path backwards from a single branch in existence at a particular time through multiple branching events ending in a single branching event for a specific experiment), then the ratio of stable $^{131}$I to unstable $^{131}$I will increase and the remaining mass of iodine will not decrease exponentially to zero.  Instead it will asymptotically approach a non-zero concentration that can be measured by any number of different techniques. 

If, as expected, this experiment fails to validate single decay-event histories, then the results of \cite{Geist01} place experimental limits on quantum kladodynamics with constant values for $\epsilon$. However, since DeWitt's picture of quantum kladodynamics is not consistent with constant $\epsilon$, we first test his picture and find that it cannot describe excited state decay.  

\section*{excited-state lifetime measurements} 

Despite metaphysical arguments\cite{Saunders01} about how EQM should assign weights to different reality branches in a superposition, the phenomenology of excited-state decay is straight forward.  To be a viable candidate, a reality-branching model must predict that the expectation value of the number of decay events observed to occur by any observer on any branch between times $t_1$ and $t_2$ is given by 
\begin{equation}
\begin{split}
\Delta N(t_2,t_1,t_0) & = N(t_0)\big[e^{-\Gamma (t_1-t_0)} - e^{-\Gamma (t_2-t_0)}\big] = N(t_0)e^{-\Gamma (t_1-t_0)}[1-e^{-\Gamma (t_2-t_1)}] \\
& =  N(t_1)e^{-(t_2-t_1)/\tau},
\end{split}
\label{Delta N(t)}
\end{equation}
where $N(t)$ is the number of identical, isolated excited particles present in a given sample at time $t$, the excited particle is created at time $t_0$, the measurement (observation) starts at time $t_1$ and stops at time $t_2$, $\Gamma = 1/\tau$ is the observed transition rate, and $\tau$ is the corresponding lifetime of the excited particle.  In CQM, Eq. \ref{Delta N(t)} is usually written with $t_1 = t_0 = 0$, but different choices for $t_0$ and $t_1$ produce different results in DeWitt-like extension of EQM. 

As a test of plausible DeWitt-like extensions of EQM, we simulate the tritium-lifetime measurements of Jones \cite{Jones1951} and Unterweger and Lucus \cite{UL} (UL) with CQM and with a few DeWitt-like extensions of EQM.  Table \ref{table2} summarizes the measurements that we simulate.   

\begin{table}[h]\begin{center}
\begin{tabular}{|c|c|c|c|l|l||} \hline 
Author & Citation      & Nominal Date    	  &      HL = ln(2)$\times\tau$	& $\tau$ (days)  \\ \hline \hline 
Jones & \cite{Jones1951}  & 1951 		  & $12.41\pm0.2$ years  & $6539\pm100$	   \\ \hline
UL & \cite{UL} & 1961, 1978, 1998-99 	  &$4504\pm9$ days 	& $6498\pm 13$       \\ \hline
AM & \cite{Akulov2004} &  $<$ 2003                & $12.296\pm0.017$ years & $6479\pm 9$  \\ \hline
\end{tabular}
\end{center}
\caption{Summary of tritium lifetime ($\tau$) measurements reported in \cite{Jones1951}, \cite{UL}, and in Akulov and Mamyrin \cite{Akulov2004} (AM) $\big($which we did not simulate for reasons described below$\big)$, including the nominal date of the measurements, where year means sidereal year. The error limits reported by Jones were probable errors (nominal 50\% coverage), those of UL were standard deviations (nominal 68\% coverage), and those of AM were probably standard deviations}
\label{table2}
\end{table}  

Prior to August 1951, Jones carried out nine counting experiments on known volumes of T$_2$ molecular gas ($98.2\pm2\%$ atomic tritium) diluted in known volumes of H$_2$ and quencher gases. He calculated the lifetime $\tau$ from
\begin{equation} 
\tau = \frac{\rho A L}{R}, 
\label{Jones}
\end{equation}
where $R = \Delta N/\Delta t$ was the average counting rate during the time period $\Delta t$, $\rho$ was the concentration of tritium in the counter at the start of the measurement, and $A$ and $L$ were the effective cross section and length of the counter.  Equation \ref{Jones} is a good approximation to Eq. \ref{Delta N(t)} because $\Delta t \le 1 \mbox{ day} << \tau$ for Jones' measurements. 

During the second half of 1998 and the first third of 1999, UL counted the number of decay events occurring in decay-event counters filled with HT derived from dilute samples of HTO dissolved in H$_2$O to calculate the massic activity $A_M = R/M$ of the samples, where $M$ is the mass of the sample.  The same type of measurement had been carried out on the same samples in 1961 by Mann, Medlock, and Yura \cite{MMY} and in 1978 by Unterweger et al. \cite{UCSM}.  The tritium in these samples was not naturally occurring, but was manufactured \cite{ULPC}, which means that it could not have been created prior to its first non-natural synthesis in 1934. \cite{Oliphant}

UL fit an exponential function of time to the three sets of massic activities to  determine the half life (lifetime times $\ln(2)$) of tritium.  Storage of the tritium as HTO in H$_2$O prevented loss of tritium by diffusion during the long storage periods between measurement periods.  

Unlike Jones and UL who recorded the decay events as they occurred, AM used a more complex mass-spectroscopy method to calculate the decay rate for tritium by comparing the ratio of $^3$He to $^4$He in ampoules containing either tritium molecules or tritium atoms (yes, they did that) and other gases just before and 568 days after sealing the ampoules. A discussion of this measurement method, much less its simulation, is significantly more complex than that of Jones and UL, so the AM results are given in Table \ref{table2} only to show that very different types of tritium decay-rate measurements all fall well within $\pm 1\%$ ($\pm65$ days) of 6500 days ($\approx 2\time10^{11}$ s).

If a DeWitt-like extension of EQM is to provide a plausible interpretation of quantum mechanics, then the results of its use in simulations of the measurements of Jones and UL must fall within approximately plus or minus one percent of $\tau = 6500$ days for all $t_2 > t_1 \ge t_0$.  In the following section we define exactly what we mean by plausible DeWitt-like extensions of EQM and use this definition to identify and simulate all such extensions for comparison with the experimental lifetimes reported by Jones and UL.  

\section*{Dewitt-like extensions of EQM}        

Here we (you the reader and I) attempt to fill in the minimum set of details not spelled out by Everett and DeWitt that are needed to simulate the decay of tritium in EQM. To start, we assume, as postulated by DeWitt, that reality branching as described by Everett's relative-state formulation of CQM does describe the cat paradox.   

According to DeWitt, the initial state of the system consists of a single branch of reality with some configuration of the following subsystems: one or more radioactive nuclei in a Geiger counter, a hammer mechanism triggered by the first Geiger count, a sealed flask of poison gas, and a live cat, all enclosed in a box.  Also according to DeWitt, after one hour, the state of the system consists of two parallel sets of branches of reality.  On each branch in one set, the original number of T nuclei are still present and the Geiger counter, hammer mechanism, flask, and cat remain in their original state.  On each branch in the other set, at least one of the T nuclei has been replaced by a $^3$He nuclei, the Geiger counter has recorded the corresponding number of counts, and the hammer mechanism, flask, and cat are in the triggered, broken, and dead state, respectively. 

Furthermore, the total wavefunction for the system will have a form in which the living cat and the dead cat and all other contents of the box are mixed in equal portions, which requires that the branches on the two sets have the same Lebesgue measure, which requires that a one to one correspondence can the established between the branches in the two sets, which is necessary so that the expectation value of the number of dead cats is 0.5 cats for all observers in the limit of infinite number of repetitions of the experiment. (For simplicity, we may refer in what follows to a set of branches as a single branch, as is consistent with Everett's description of his formulation, and only spell out all of the details when necessary for clarity.) 

Clearly the total wavefunction for this system is a very complex mathematical expression, and the total amount of information provided by DeWitt and Everett is nowhere near complete enough to guide us to write out this wavefunction in sufficient detail to use it as a guide to simulate the cat paradox.  For instance, which portions of this apparatus meet Everett's definition of observers as automatically functioning machines, possessing sensory apparatus and coupled to recording devices capable of registering past sensory data and machine configurations.  The dead cat, the broken flask, and the final state of the hammer all seem to register past sensory data. The cat, flask, hammer, and Geiger counter, all seem to possess sensory apparatus.  Is only one of these the observer, or are all observers?  If the later, what exactly does each observe, the death of the cat, or just the preceding stage in the whole process.  And finally, what does the total wavefunction look like in terms of the wavefunctions of the individual subsystems in terms of sums and products, etc?      

Luckily, we do not need to answer these question because Everett states that  
\begin{quote}
\textit{... a wave function that obeys a linear wave equation everywhere and at all times supplies a complete mathematical model for every isolated physical system without exception. It further postulates that every system that is subject to external observation can be regarded as part of a larger isolated system.}
\end{quote}
Therefore, we do not need to work with a wavefunction that mixes a living cat and a dead cat in equal portions at some time.  Instead, we only need to work with a wavefunction that mixes a tritium nucleus T with a  helium-3 nucleus $^3$He in equal portions at some time.    

Next suppose that a single tritium nucleus, which was created at time $t_0$, exists on a single branch of reality at time $t_1$ inside a Geiger counter/computer (GCC) that detects and records the time that Geiger-counter pulses are detected. Let $\phi_T(t)$ be the wavefunction that describes this nucleus. Let $\psi^0[...]$ be the GCC wavefunction. In this case, Eq. 10 of \cite{Everett01} can be written as
\begin{equation} \label{DeWitt1}
 \psi^{S+0}(t_1) =  \exp(-i E_1 t_1/\hbar) \phi_1 \psi^0[...] 
\end{equation}
to describe the initial state of the tritium nucleus-GCC system in a more application-specific notation that explicitly includes time since time is key to tritium lifetime measurements, where $\phi_1$ is the normalized, excited eigenstate with respect to the weak force that we call a T nucleus, and $\phi_0$ is the normalized ground state that we call a $^3$He nucleus.      

We still need to know how to write a wavefunction that has a form in which a T nucleus and a $^3$He nucleus are mixed in equal portions in EQM.  In this connection, Everett states, 
\begin{quote}
\textit{
We therefore seek a general scheme to assign a measure to the elements of a superposition of orthogonal
states We require a positive function $m$ of the complex coefficients of the elements of the superposition, so that $m(a_i)$ shall be the measure assigned to $\phi_i$. ... in other words, we have $m(a_i) = a_i^* a_i$ }
\end{quote} 
We will define $\epsilon$ in \cite{Geist01} with $m(a_1)$ below. 
 
If no decay event has been detected between times $t_1$ and $t_{1/2}$, then according to DeWitt, the T and $^3$He nuclei are mixed at time $t_{1/2} = \ln(2) \tau$ in equal portion given by 
\begin{equation} \label{DeWitt2}
\begin{split}
\psi'^{S+0}(t_{1/2}) = \psi^{S+0'}(t_{1/2}) 
= & \sqrt{1/2} \; \exp(-i E_1 t_{1/2}/\hbar)\; \phi_1\;\psi^0[...] \\		    
+ & \sqrt{1/2} \; \exp(-i E_0 t_{1/2}/\hbar)\; \phi_0\;\psi^0[...]. 
\end{split}
\end{equation}

The first and second terms on the left side of Eq. \ref{DeWitt2} are in the notation used in Eqs. (12) and (18) of \cite{Everett01}, respectively, and $E_0$ and $E_1$ are the energy eigenvalues of the T and $^3$He nuclear eigenstates, respectively.     

However, if a decay event has occurred at time $t_d$, where $t_1 \le t_d < t_{1/2}$, then to fit Everett's definition of the final state of a superposition following an observation, the T and $^3$He nuclei eigenstates must be mixed in portions given by 
\begin{equation} \label{DeWitt3}
\begin{split}
\psi'^{S+0}(t_d) = \psi^{S+0'}(t_d) 
   =  a(t_d) \; & \exp(-i E_1 t_d/\hbar) \; \phi_{1} \; \psi^0[...E_1] \\
+ \sqrt{1-a*(t_d)a(t_d)} \; & \exp(-i E_0 t_d/\hbar) \; \phi_{0} \; \psi^0[...E_0],
\end{split}  
\end{equation}
where $a(t_d)$ is a complex number with magnitude between 0 and 1. 
 
Next we examine Eq. \ref{DeWitt2} with respect to Everett's definition of a good observation of a property of an eigenfunction:   
\begin{quote}
\textit{A good observation of a quantity A, with eigenfunctions $\phi_i$, for a system S, by an observer whose initial state is $\psi^{0}$, consists of an interaction which, in a specified period of time, transforms each (total) state 
\begin{equation} 
\psi^{S+0}(t) =  \phi_i \psi^0[...],
\end{equation}
into a new state
\begin{equation} 
\psi^{S+0'}(t) =  \phi_i \psi^0[...\alpha_i],
\end{equation}
where $\alpha_i$, characterizes$^7$ the state $\phi_i$. (The symbol, $\alpha_i$, might stand for a recording of the eigenvalue, for example.) That is, we require that the system state, if it
is an eigenstate, shall be unchanged, and (2) that the observer state shall change so as to describe an observer that is "aware" of which eigenfunction it is; that is, some property is recorded in the memory of the observer which characterizes $\phi_i$, such as the eigenvalue.}
\end{quote} 

Because $\phi_1$ in Eq. \ref{DeWitt1} is an eigenstate, the quantity $\psi^{S+0'}(t_{1/2})$ in Eq. \ref{DeWitt2} should be $\phi_1 \psi^0[...]$ without the second term on the right-hand side of that equation. Probably, EQM can be reformulated in a way that removes this problem.  Until that is done, EQM cannot describe excited-state decay, which has been ubiquitous in CQM from its inception, and we must reject Everett's claim that EQM is a \lq\lq complete theory".   

DeWitt either ignored this problem or solved it by implicitly generalizing Everett's formulation of EQM such that Everett's definition of a good observation only applied to ground states with respect each fundamental force.  Whether or not this change can be made consistent with the remainder of Everett's formulation is beyond the scope of this report.  Assuming this to be the case, we describe the T-$^3$He nuclear system for time $t$ such that $t_1 < t < t_2$ by
\begin{equation} \label{DeWitt4}
\begin{split}
 \psi^{S+0'}(t) =  a_1(t) & \exp(-i E_1 t/\hbar) \; \phi_{1} \psi^0[\alpha_1] \\
	+ \sqrt{1-a_1(t)*a_1(t)} & \exp(-i E_0 t/\hbar) \; \phi_{0} \psi^0[\alpha_0],
\end{split}  
\end{equation}
where $\alpha_1 = 0$, $\alpha_0 = 0$ if one or more decay events are detected and recorded by the GCC between times $t_1$ and $t_2$, and $\alpha_0 = 1$ if no decay events are recorded by the GCC between times $t_1$ and $t_2$. Even though $\alpha_i$ is not an eigenvalue, it still satisfies Everett's requirement that the symbol, $\alpha_i$ change so as to describe an observer that is "aware" of which eigenfunction it is ..." Also, for use in simulating the detection and counting process, let $t_d$ be the time when the first count was recorded if one or more decay events are detected and recorded by the GCC between times $t_1$ and $t_2$  

Time dependence of the amplitudes $a_i$ of the eigenstates in a superposition during the transformation of the wavefunction from its initial state to its final state is crucial to the continuous evolution of the wavefunction.  It is only after \lq\lq a specified period of time" following the interaction that the amplitudes become independent of time in his formulation.  

It appears that the only aspect of Eq. \ref{DeWitt4} that violates Everett's formulation is the transformation of the excited eigenstate $\phi_1$ into a superposition rather than the same eigenstate. But, this violation is crucial to DeWitt's picture of the cat paradox.     

Next, we test Eq. \ref{DeWitt4} against DeWitt's description of the cat paradox by changing a few words to describe tritium decay. Our additions are in bold face, words replaced are marked with strike outs, and extraneous words deleted.  The result is, 
\begin{quote}\textit{... The counter contains a trace of radioactive material - just enough \textbf{(one molecule containing a tritium nucleus)} that in \sout{one hour} \textbf{12.3 years} there is a 50\% chance one of the nuclei will decay ... and therefore an equal chance that \sout{the cat will be poisoned} \textbf{a $^3$He atom will be present}.  \sout{At the end of the hour} \textbf{After $t_2 = 12.3$ years, which is the half-life of tritium $t_{1/2}$,} the total wavefunction for the system will have a form in which the \sout{living cat and the dead cat} \textbf{tritium nucleus and an $^3$He nucleus} are mixed in equal portions.  Schroedinger felt that the wave mechanics that led to this paradox presented an unacceptable description of reality. However, Everett, Wheeler, and Graham's interpretation of quantum mechanics pictures the \sout{cats} \textbf{nuclei} as inhabiting two simultaneous, noninteracting, but equally real worlds.}\end{quote}     

While requiring a modification of EQM as formulated, it is clear that Eq. \ref{DeWitt4} condenses DeWitt's picture of the EQM description of the cat paradox to an EQM description of the radioactive decay of tritium.  The only thing missing is the actual functional form of $\epsilon(t) = m\big(a_1(t)\big) = a^*(t)a(t)$, which, as defined previously, is the measure of the branch on which no decay event was detected, which is also the apparent probability with post-split observers are assigned to that branch immediately following a decay event. . 

DeWitt provided two constraints on $a_1(t)$: 1) $a_1(t_1) = 1$, and 2) $a_0(t_{1/2}) = 0$ if no decay event occurred between times $t_1$ and $t_{1/2}$.  Everett provided a third constraint: $a_0(t) = 1$ if a decay event occurred at time $t_d$, $t_1 \le t_d < t_2$, and $t \ge t_d$.  

To start, we restrict our search to monotonically decreasing functions of time, and note that 
\begin{equation} \label{epsilon}
\epsilon(t) = a_1^*(t)a_1(t) = \exp\big((-t-t_x)/\tau\big)
\end{equation}
satisfies the constraints set by Everett and DeWitt with time $t_x$ unspecified. Accepting that constraints on $t_x$ based on CQM might not apply in EQM, we allow that $t_x$ might be $t_0$, $t_1$, or $t_d$ depending upon the experiment being simulated. 

One virtue of Eq. \ref{epsilon} compared to other monotonic functions that satisfy the constraints is that it requires no fundamental constants beyond those already in the standard model, which is important if EQM is to be no more than a reformulation of CQM rather than an extension of the standard model. It would not be surprising if Eq. \ref{epsilon} is what DeWitt had in mind.    

What happens on the branch on which no decay event occurred? Are further decay events forbidden or allowed? Conveniently, the probability of any tritium nucleus undergoing two decay events in DeWitt-like extensions of EQM during the measurement interval $t_2-t_1$ used by \cite{Jones1951} and \cite{UL} is negligible in comparison to the already small probability of any single particle undergoing a decay event during that time.  In other words, the simulation results obtained for both forbidden and allowed multiple decay events per single excited particle in the experiments described by Jones and UL are identical for all practical purposes.  Therefore, we simulate only the former knowing that our results describe both possibilities. 

A final issue that must be resolved before the simulations can be carried out is that the measurements are simultaneously carried out on a great number of approximately isolated (with respect to the weak force) nuclei, so the simulations should be carried out the same way.  An ideal measurement would perfectly isolated each nucleus. However, the tritium nucleus in an atom or molecule is well isolated by its, small size and the nature of the weak force, as well as the electron cloud.  In fact, it is so well isolated that the lifetime of a bare tritium nucleus (triton) differs from that of a triton in an atom or molecule by much less than 1\%. \cite{Akulov2004}. 

We still must show that branching does not add errors other than that described by $\tau_A$ and $\tau_B$ to the results obtained by an observer meeting Everett's definition of an observer measuring the lifetime of tritium in DeWitt's picture of radioactive decay. We do this next. 

\section*{Inside and outside views of reality branching}

At this point, it is useful to consider an excerpt from Tegmark's\cite{Tegmark} explanation of the difference between the physics of EQM reality branching and the experience of an observer during EQM reality branching: (Note I don't agree with some of his assertions, but do think it is a good way to think about EQM.)
\begin{quote}
\textit{Everett's many-worlds interpretation has been boggling minds inside and outside physics for more than four decades. But the theory becomes easier to grasp when one distinguishes between two ways of viewing a physical theory: the outside view of a physicist studying its mathematical equations, like a bird surveying a landscape from high above it, and the inside view of an observer living in the world described by the equations, like a frog living in the landscape surveyed by the bird.}
\end{quote}

To test the physical viability of reality branching, we must simulate the observations of a typical observer (inside view) attempting to measure the lifetime of a sample containing a large number tritium atoms starting with a single observer on a single branch at time $t_0$ and ending with multiple observers on multiple branches at time $t_3$.  Figures \ref{branches} \textbf{a} and \textbf{b} illustrate the outside and inside views in black and red, respectively.   

To avoid unnecessary complexity, we simulate the ideal case of an observer observing decay events of isolated tritium atoms \textsf{A, B}, and \textsf{C} consecutively by selecting a new atom and resetting $t_0, t_1, t_2$, and $t_3$ after each observation.

As described above, this is not how Jones and UL carried out their measurements.  Instead, due to time limitations, they used macroscopic samples containing macroscopic numbers of tritium atoms and counted the number of decay events occurring over times of the order of hours to days, to provide approximations to the results that would be obtained with serial measurements of the same number of perfectly isolated tritium atoms. 

Importantly, Akulov and Mamyrin \cite{Akulov2004} measured the effect of imperfect isolation. They report a difference of approximately $0.26\%$ between the lifetime of tritium atoms in the gas phase and that of a tritium atoms chemically bonded in a molecule.  Therefore the effect of imperfect isolation in HTO is negligible at the level of a few percent, and room-temperature collisions of tritium containing molecules with other molecules as well as with other tritium containing molecules are completely negligible at this level of accuracy.   

\subsection*{Outside view}

\begin{figure}[h] 
  \centering
\includegraphics[width=10cm,height=7.6cm,keepaspectratio]{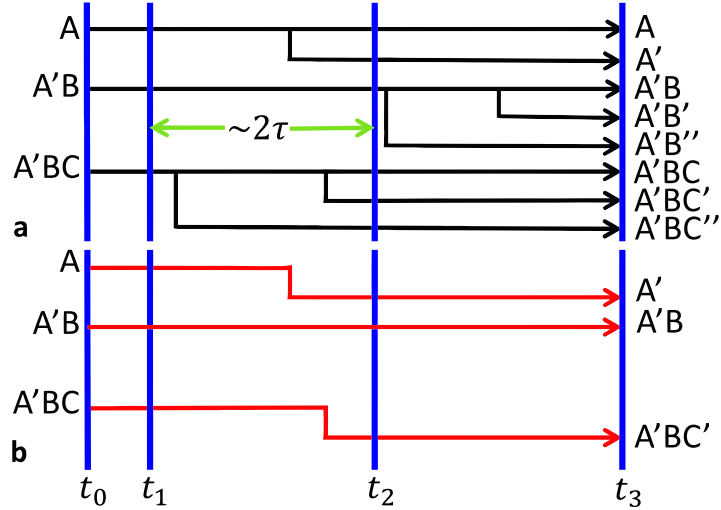} 
\caption{ Illustration of the outside (\textbf{a}) and inside (\textbf{b}) views of one of the DeWitt-like extension of EQM that corresponds to Eq. \ref{DeWitt2} for serial observations over a fixed time period from $t_1$ to $t_2$ of the decay (or lack thereof) of three tritium atoms, where time $t_0$ is reset to 0 at the beginning of each observation. This figure is compressed for clarity as described below.} 
\label{branches}
\end{figure}

The horizontal black lines in Fig. \ref{branches}\textbf{a} are branches that the particles \textsf{A, B}, and \textsf{C} follow through the branching realities. The vertical black lines represent decay events.  Two horizontal black lines follow each decay event to represent reality branching into a branch on which the decay event did not occur and a branch on which the decay event did occur.    

From the outside view, the single observer present at time $t_0$ prepares a measurement such that particle \textsf{A} is inside a decay-event counter in an excited state at time $t_0$. The observer turns the counter on and sets it to zero at time $t_1$, turns it off at time $t_2$, and records the counter output.  The vertical black line between the blue lines at times $t_1$ and $t_2$ indicates the occurrence of a decay event while the counter was active.  

At time $t_3$ the original branch \textsf{A} has branched into two branches \textsf{A} and \textsf{A'}.  On branch \textsf{A}, no decay event has occurred, and the version of the original observer on that branch has not recorded a decay event. On branch \textsf{A'}, a decay event has occurred while the counter was active, and the version of the original observer on this branch has recorded a decay event. 

Having completed the first measurement, both versions of the original observer prepare measurements and reset their timers to $t_0$ such that their versions of particle \textsf{B} are inside a decay-event counter in an excited state at time $t_0$. Both versions of the original observer turn their counters on and set them to zero at time $t_1$, turn them off at time $t_2$, and tally the result before time $t_3$.  For simplicity we only show one of these two branches in Figure \ref{branches}\textbf{a}.  It shows the branch \textsf{A'B} for observer \textsf{A'} branching into three branches \textsf{A'B}, \textsf{A'B'}, and \textsf{A'B''} because two decay events occur on branch \textsf{A'B} between times $t_1$ and $t_3$.  But neither of these decay events are recorded on any branch because they occur after the counters have been turned off at time $t_2$.  

Having completed the second measurement, all three versions of the version of the original observer that we followed prepare measurements and reset their timers to $t_0$ such that their versions of particle \textsf{C} are inside a decay-event counter in an excited state at time $t_0$. All three versions of the original observer turn their counters on at time $t_1$, turn them off at time $t_2$, and tally the result before time $t_3$.  Again, Figure \ref{branches}\textbf{a} only shows one of the branches created during the second measurement. It shows the branch \textsf{A'BC} for the observer present on \textsf{A'B} at time $t_3$ branching into three branches \textsf{A'BC}, \textsf{A'BC'}, and \textsf{A'BC''} because two decay events occur on branch \textsf{A'B} between times $t_1$ and $t_2$.  Therefore the observers on branch \textsf{ABC'} and the observer on branch \textsf{ABC''} each recorded a decay event because those decay events occurred during the time when the counters were active.  

Note that the description given above describes serial measurements of the decay of excited-state particles \textsf{A, B}, \textsf{C}, whereas Fig. \ref{branches}\textbf{a}, which has been compressed for clarity, looks like it describes parallel measurements, but this is not the case.  Actually branch \textsf{A'B} starts at the right hand side of branch \textsf{A'} and branch \textsf{A'BC} starts at the right hand side of branch \textsf{A'B}. The full figure would also include branches \textsf{ABC}, \textsf{ABC'}, \textsf{ABC''}, \textsf{AB'C}, \textsf{AB'C'}, \textsf{AB'C''}, \textsf{AB''C}, \textsf{AB''C'}, \textsf{AB''C''}, etc. 

A similar figure for parallel measurements of the simultaneous decay of particles \textsf{A, B}, and \textsf{C} is much more complex because each decay event on each branch causes branching to occur at the same time on every other branch, not only generating new branches on every previously existing branch with each new branching event, but also generating new universes with each new branching event. Even this does not describe the full complexity of the outside view because different branches have different Lebesgue-measure weights according to Everett.  Does this mean that there are many possibly parallel ground state branches and many parallel excited-state branches each with and its own version of the original observer, none of which are aware of the other.  Or are there only two branches and multiple sub-branches?  Luckily, these issues do not arise in the inside view, where we know that each observer is only aware of its own environment.  
  
\subsection*{Inside view}

Now consider the inside view of the same decay events. The horizontal red lines in Fig. \ref{branches}\textbf{b} are branches that the particles \textsf{A, B}, and \textsf{C} follow through the branching realities. The vertical red lines represent decay events that are detected and recorded by the observer.             

By definition, the inside view can only follow a single observer along a single branch at each decay event.  Not only does the observer have no control over which branch is followed, the observer is not even aware of the branching according to Everett. To simulate the results of a typical observer, we must choose to follow one branch or another by a random process.  

The single observer present at time $t_0$ prepares a measurement such that particle \textsf{A} is in an excited state inside a decay-event counter at time $t_0$. The observer turns the counter on at time $t_1$, turns it off at time $t_2$, and tallies the result at time $t_3$.  The vertical red line between the blue lines at times $t_1$ and $t_2$ indicates that the observer detected and recorded the occurrence of a decay event while the counter was active. For the case illustrated in Fig. \ref{branches}\textbf{b}, we follow branch \textsf{A'} of the outside view, but could just as well chosen to follow branch \textsf{A'B}. 

Now that the first measurement is complete, the version of the original observer that we are following prepares a measurement such that its version of particle \textsf{B} is in a decay-event counter in an excited state at time $t_0$. This time we follow the version of the original observer who at time $t_3$ did not observe a decay event because none occurred while the event counter was active and there is no second branch. But this observer has no way of knowing about the existence or non-existence of other branches nor whether any decay events occurred on other branches.    

Now that the second measurement is complete, the version of the original observer that we are following prepares a measurement such that its version of particle \textsf{C} is in a decay-event counter in an excited state at time $t_0$. Again, based on a random process, we follow one of the three versions of the original observer that exist in the outside view at time $t_3$.  This time, two decay events occur between times $t_1$ and $t_2$ while the counter is operating, but the observer that we follow detects and records only one decay event because from the inside view there is only a single branch.  

If the observations are carried out consecutively as describe earlier, the single branch is constructed by putting the \textsf{A'B} branch at the end of \textsf{A'} branch, and the \textsf{A'BC} branch at the end of \textsf{A'B'} 
branch in accordance with the outside view.  

If the observations are carried out simultaneously, the labels in Fig. \ref{branches}\textsf{b} are replaced with the particle names \textsf{A, B}, and \textsf{C}, on both the left-hand side and the right-hand side of the figure.  But the branches are identical for both serial and parallel measurements in accordance with experimental fact. This one-to-one mapping of the inside view of simultaneous and serial lifetime measures demonstrates that the results of real decay-event counting experiments that are carried out simultaneously over a known time period can be simulated as consecutive decay-event counting experiments, each carried out over time intervals of the same duration as the time period used in the real measurement.  

In fact, as already pointed out, decay-event counting experiments that are carried out simultaneously are done under conditions that approximate isolated particles, and if corrections are made then they are based on theoretical calculations of small effects or extrapolation of results collected at different level of isolation to the condition of complete isolation\cite{Schmoranzer, Volz01}.  This, of course, is the result that would be obtained in a serial experiment that measured the decay time of the same sample of isolated particles one at a time.  It is the requirement to record enough counts to produce low statistical uncertainties that precludes serial measurements for long-lifetime particles.   

At this point, it is perhaps of interest to continue Tegmark's\cite{Tegmark} explanation of the difference between the physics of EQM reality branching and the experience of an observer during EQM reality branching: 

\begin{quote}
\textit{From the bird perspective, the Level III multiverse is simple. There is only one wave function. It evolves smoothly and deterministically over time without any kind of splitting or parallelism. The abstract quantum world described by this evolving wave function contains within it a vast number of parallel classical story lines, continuously splitting and merging, as well as a number of quantum phenomena that lack a classical description. From their frog perspective, observers perceive only a tiny fraction of this full reality. They can view their own Level I universe, but a process called decoherence--which mimics wave function collapse while preserving unitarity--prevents them from seeing Level III parallel copies of themselves.}
\end{quote} 

If the outside view is simple, why didn't Tegmark illustrate it for us?  What does it look like, this wavefunction that evolves smoothly and deterministically over time without any kind of splitting or parallelism and yet describes a vast number of parallel classical story lines?  What does this wave function look like for the simple three-particle experiment illustrated in Fig. \ref{branches}, where the outside view is clearly very complex?  If Tegmark means that the bird is myopic and only sees a blur that shows no hint of splitting and branching, then this outside view is of no use to a physicist trying to predict what human observers actually measure in any given experiment. 

Fortunately, we don't need the outside view of the wave equation to simulate the tritium lifetime measurements of Jones and UL with EQM.  Instead, we test different possibilities for $\epsilon(t)$ in Eq. \ref{epsilon} and compare the results of the simulation with the reported results.  Without $\epsilon(t)$, there is no EQM.  When $\epsilon(t) = 0$, EQM is CQM, and reality branching does not occur.  EQM with reality branching is not even possible unless there exists a mathematical expression for $\epsilon(t)$ such that simulations of the tritium-lifetime measurements of Jones and UL produce results consistent with the results reported by these authors within plus or minus a few percent.  

\section*{Simulation procedure} 

\begin{figure}[htbp] 
\includegraphics[width=17.5cm,height=13.3cm,keepaspectratio]{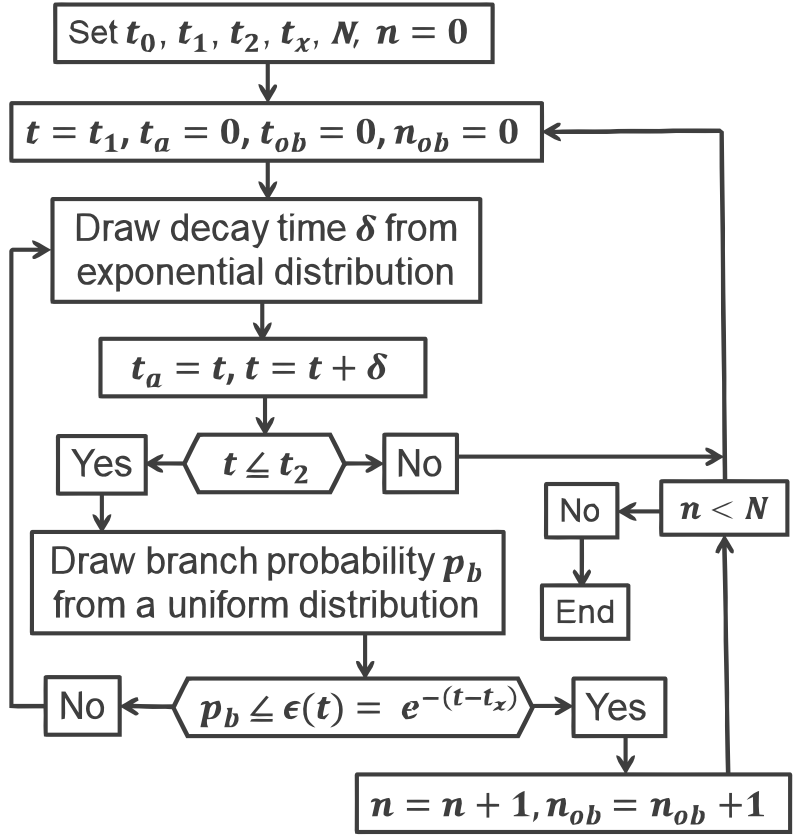}
\caption{\small Flow chart describing simulations of excited-state decay-rate measurements. Time is normalized to the observed lifetime $\tau$ of the excited particle. The quantities $t_0$ and $t_1$ are the respective times from the creation of the excited particle and from the start of a decay-rate experiment, $t_2$ is the time when the measurement ends, $N$ is the number of independent measurements carried out, and $n$ is the measurement number. The quantity $\delta$ is the random time between decay events in the outside view, $t$ is the time when a decay event occurs in the outside view, $t_a$ is $t_1$ at the start of a new measurement and is set to the time in the outside view when the antecedent (previous) decay event occurred following each decay event that occurs during the measurement period. The quantity $t_{ob}$ is the time when a decay event is observed (detected and recorded) by the randomly-chosen, ever-branching observer whose history we simulate and $n_{ob}$ counts the number of decay event observed (detected and recorded) between times $t_1$ and $t_2$, which is either zero or one. The quantity $p_b$ is the branch probability, which is compared to $\epsilon$ to determine on which randomly chosen branch the observer that we follow ends up, $t_x$ is $t_0$, $t_1$, or $t_a$ depending upon which DeWitt extension of EQM we simulate, and $\epsilon$ from Eq. \ref{epsilon} is the probability of the randomly chosen observer remaining on the excited state branch following a decay event, where $t_x$ has the different values given in Table \ref{Table02} below for different DeWitt-like decay models. } 
\label{FlowChart}
\end{figure}

To test the physical viability of reality branching, we simulate the history of a single observer performing a large number $N$ of measurements of the lifetime $\tau$ of a single isolated excited particle, one measurement at a time. Each measurement ends either when a decay event is detected or at the end a measurement period, which ever comers first. Time is measured in units of $\tau$ for simplicity. 
We randomly follow a single continuous branch in time through the branching realities.  Figure \ref{FlowChart} shows a flow chart for this branch in a DeWitt-like extension of EQM. 
 
To start each simulation, a number of parameters are set to their initial values as described in Fig. \ref{FlowChart}, the simulated decay time $\delta$ is drawn from an exponential distribution by calculating
\begin{equation}
\delta = -\ln(u)
\label{-ln(u)}
\end{equation}
where $u$ is drawn from a uniform random distribution $U$, $0 \le U < 1$, and $p_b$ in Fig. \ref{FlowChart} is drawn from the same uniform random distribution given by 
\begin{equation}
\epsilon_i = \exp(-\Delta t_i) =  \exp\big(-(t-t_x)/\tau\big), 
\end{equation} 
where $\tau$ is the experimentally observed value $\tau_{ob}$ and $t_x$ is different for different DeWitt-like excited-state decay models. 

After each decay event, there will be at least one observer on a ground-state branch of reality who has detected a decay event, incremented $n_{ob}$, and recorded the time $t_{ob}$ the decay event occurred.  Similarly there will also be at least one observer on an excited-state branch still waiting to observe a decay event. The value of $p_b$ relative to $\epsilon_i$ randomly determines whether we follow an observer on a ground-state branch or an observer on an excited-state branch through the branching realities.  After $N$ decay events have been detected by the observer that we are following, there will be $N$ branch points on the branch that we followed. 

The process in Fig. \ref{FlowChart} does not simulate all of the reality branches generated in the outside view during the parallel measurements of the of the number of decay events occurring in samples containing a large number of tritium atoms of the type described by Jones and by UL.  But it does provide a simulation of the reality branches that Jones and UL would experience in their measurements if one of DeWitt's extensions of EQM is correct and they carried out their measurements one particle at a time.  Importantly, as shown above, there is a one-to-one mapping of the branches experienced by a simulated observer performing measurements in parallel onto the branches experienced by a simulated observer performing the measurements serially because, at each branching event, the observer that we follow is aware at most of only one such event whether the observer is performing the observations in parallel or serially.      

\section*{Simulations of DeWitt-like extensions of EQM} 

Four different tritium lifetime determination models are listed in Table \ref{Table02}. Model 0 predicts the number of counts that Jones would have reported in the absence of reality branching.  Models 1 and 2 respectively predict the number of counts that Jones would have reported for two different DeWitt-like models of $\epsilon(t)$ under the same experimental conditions.  These numbers correspond to lifetimes of over 2 million years. The results of Model 3, which are more complex because they depend upon $t_0$, are not given in the table, but are given in the section describing the lifetime determination reported by UL \cite{UL}.  
  
\begin{table}[h] \label{Models}
\begin{tabular}{|c|c|c|c|} \hline 
Model Number & Model Type & Model & Decay Events  \\ \hline \hline 
0 & No branching  & $\epsilon(t) = 0$  & $385,000 \pm 1,200$  \\ \hline
1 & DeWitt-like &  $\epsilon(t) = \exp\big(-(t-t_a)/\tau \big)$ & $31\pm 11$ \\ \hline
2 & DeWitt-like &  $\epsilon(t) = \exp\big(-(t-t_1)/\tau \big)$ & $24\pm 10$ \\ \hline
3 & DeWitt-like &  $\epsilon(t) = \exp\big(-(t-t_0)/\tau \big)$ & Depend on $t_0$. See Fig. \ref{ULsim}\\ \hline
\end{tabular}
\caption{Results of simulations of tritium lifetime measurements of Jones\cite{Jones1951} for $\epsilon(t) = 0$ (no reality branching) and two different models of $\epsilon(t)$ from Eq. \ref{epsilon}. The parameters $\delta, t_a, t_0, t_1$, and $t_2$ are defined in the caption of Fig. \ref{FlowChart}. The uncertainty is the 95\% confidence interval for the number of counts recorded.} 
\label{Table02}
\end{table} 

\subsection*{Simulation predictions of tritium lifetime reported in Ref. \cite{Jones1951}}

Jones\cite{Jones1951} reported the average of 17 independent measurements of the decay rate of an accurately diluted samples of 98.2\%-pure, molecular-tritium gas in hydrogen, further accurately diluted in a quencher gas to give an average counting rate of approximately 45 counts per second in different measurements in different decay-event counters. With the measured lifetime of 6500 days, this corresponds to approximately $2.5 \times 10^{10}$ tritium atoms for each measurement. 

Jones did not report the duration of each measurement, but did report that no drift was observed in a preliminary measurement that lasted two days, and used this fact to justify no correction for drift.  Therefore, Jones' measurements probably took no longer than two days.  

With different models of $\epsilon(t)$ from Eq. \ref{epsilon}, we simulated the number of decay events that Jones would have detected if
\begin{itemize}
  \item the lifetime of tritium was 6500 days, 
  \item the tritium sample contained $N = 2.5\times10^{10}$ tritium atoms, 
  \item each measurements started at time $t = t_1$, 
  \item each measurement stopped when $t > t_2-t_1 = 0.000154 \tau \approx 1$ day, or when a decay event was detected. 
\end{itemize} 

 As mentioned previously, these results apply to Models 1 and 2 with multiple decay events allowed on the persistent excited-state branch and with no more than one decay event allowed on the excited-state branch because the probability of a second decay event occurring between times $t_1$ and $t_2$ is negligible compared to that for a single event. 

Clearly both versions of Models 1 and 2 must be ruled out as incapable of describing the experimental facts of the radioactive decay of tritium.  The reason that these models give such poor results is that the measurement time $t_2-t_1$ is very short compared to the lifetime $\tau$ of tritium and $\epsilon\big((t-t_x)/\tau\big) = 1$ at the beginning of each measurement and decreases exponentially to near zero over many lifetimes, whereas with Model 0, $\epsilon\big((t-t_x)/\tau\big) = 0$ for all $t > t_1$. 

These results probably could have been derived analytically from the equations in Table \ref{Table02}.  However, considering the \lq\lq mind-boggling" nature of \lq\lq Many Worlds" theories, it seems that very straight forward descriptions and simulations of the actual decay process might be more understandable and persuasive than three-step or four-step analytic calculations. 

Recall that there is still a third DeWitt-like model that must be considered. This model, which is  Model 3 in Table \ref{Models} is a random process with a memory, namely
\begin{equation}
\epsilon(t) = \exp\big(-(t-t_0)/\tau \big),
\label{Model 3}
\end{equation}
where $t_0$ is the time when the excited state is created. While this model is obviously wrong in CQM, who knows what might work in DeWitt-like versions of EQM? 

Jones did not report the source of the 98.2\%-pure molecular tritium gas that was the starting point for his measurement.  If the tritium was manufactured, then $t_0$ could not be earlier than 1934 because that was when it was first manufactured.\cite{TritiumDiscovery} If it was from a natural source, there is no hard lower bound, only a soft lower bound determined by a factor of two decrease in concentration with the duration of each period of time equal to a half life since the creation of the tritium nuclei.  Therefore, instead of Jones' results, we compare the results of a different set of measurements to simulations of the counting rate of tritium samples as a function of $t_1-t_0$.      

\subsection*{Simulation predictions of tritium lifetime reported in Ref. \cite{UL}}

\begin{figure}[h] 
\centering
\includegraphics[width=10cm,height=7.6cm,keepaspectratio]{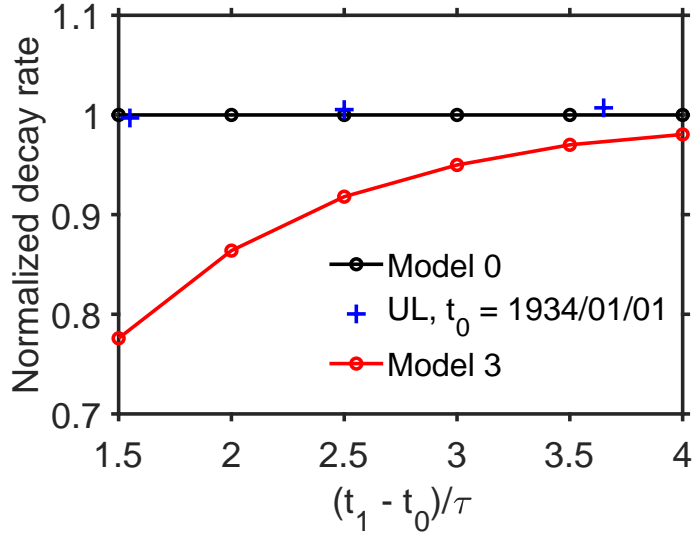} 
\caption{Black curve and Red curve - Normalized decay rates for Model 0 (no reality branching) and Model 3 (Eq. \ref{Model 3}) as a function of the time interval $t_1-t_0$ between the time when the tritium was manufactured and the time when its massic activity was measured, respectively.  Blue crosses - Normalized decay rate reported by Unterweger and Lucas\cite{UL} under the assumption that $t_0$ is January 1, 1934.} 
\label{ULsim}
\end{figure}

Unterweger and Lucas (UL) \cite{UL} described measurements of the Massic activity of manu-factured\cite{ULPC} tritium in tritiated water carried out at NBS/NIST at three different times over 38 years. Figure \ref{ULsim} shows the results of simulating tritium decay-rate measurements as a function of the time $t_0$ from the creation of the tritium nuclei until the time $t_1$ at the start of a decay-rate measurement for $\epsilon(t)$ from Model 3 in Eq. \ref{Model 3}.   

UL did not determine the lifetime from the decay rate of known quantities tritium, but by fitting Eq. \ref{Delta N(t)} to the Massic activity of the same samples of tritium measured at the three different times. They did not report the duration of their measurements, but did report different dates for each of measurement, so the measurements probably lasted less than one day. Based on this fact we again set the measurement time to $t_2-t_1 = 0.000154\tau \approx$ one day and again set the total number of measurements simulated to $N = 2.5\times10^{10}$ to minimize statistical errors. 

All of the data in Fig. \ref{ULsim} are normalized to Eq. \ref{Delta N(t)} with $\tau = 6500$ days.  The blue crosses are the results that UL reported plotted under the assumption that $t_0$ is January 1, 1934. If $t_0$ is a later date, then the blue crosses are displaced to the left in the graph by the same number of lifetimes.  The results reported by UL agree with the predictions of Model 0 within $\pm$1\% for all three values of $t_1-t_0$.  On the other hand, if Model 3 provided an accurate description of tritium decay, then the UL results (blue crosses) would fall within a few percent of the red line at the same values of $t_1-t_0$. Clearly, Model 3 must be ruled out as a viable extension of EQM to excited-state decay for both single and multiple decays allowed along the persistent excited-state branch.   

The results described here definitively rule out DeWitt's picture of EQM with a specific model for $\epsilon(t)$ to describe serial measurements of the lifetime of single radioactive nuclei.  Therefore, this version of EQM is not a viable basis for a universal wavefunction and all of the predictions including reality branching based on it are subject to doubt. Below we shown that reality branching can describe decay of the lowest excited state if, and only if, $\epsilon(t)$ is constant, which rules out all other DeWitt-like extensions of EQM.   

Note that debunking DeWitt's version of EQM does not invalidate the Weisskopf and Wigner (WW) \cite{WW} derivation of the line shape of an excited-state transition.  This model produces a very useful result.  However, the WW model is not a continuous, deterministic change of state of an isolated nucleus, atom, or molecule with time according to a wave equation $\partial \psi/\partial t = A \psi$, where $A$ is a linear operator because the energy lost during the transition does not leak out over time as built into the model, but is released all at once at a random time.  What the WW model describes very well is the expectation value of the spectrum of electronic-state decay in the limit as the number of particles in the sample being measured increases without bound.  

However, debunking DeWitt's version of EQM has only marginal relevance to the question of whether or not Everett's formulation of EQM can describe excited-state decay, much less is actually a reliable description of reality, because it violates Everett's formulation.  Therefore, the question of whether or not Everett's formulation can be extended to describe excited-state decay in a way that preserves reality-branching is still unanswered. We address this question next. 

\section*{Constant $\epsilon$ extensions of EQM}    

As mentioned above, the only other extension of EQM to excited-state decay that I know of is in \cite{Geist01}. This study did not simulate the results of any experiment.  Instead it analytically calculated the expectation value of the lifetime determined by a single observer recording decays of a single particle for the situation where multiple decay events are allowed on the excited state branch in the limit of an infinite observation time. 

The result for this case was given by
\begin{equation} \label{AB}
\tau_B = (1- \epsilon) \tau_A,
\end{equation} 
where $\tau_B$ is the expectation value of the lifetime of the excited state in the outside view and $\tau_A$ ($\tau$ here) is the expectation value of the lifetime as measured by an observer (inside view).  This result is important because the mathematical steps used to derive it are reversible.  Specifically, $\tau_B = (1- \epsilon) \tau_A$, if and only if $0 < \epsilon < 1$, where $\epsilon$ is the ratio immediately following a decay event of the number of observers who do not observe a decay event to the total number of observers who are generated from the original observer on a single branch immediately following a decay event.  This rules out all DeWitt-like versions of EQM, but does not rule out EQM.  

\begin{figure}[htbp] 
  \centering
\includegraphics[width=20cm,height=15.2cm,keepaspectratio]{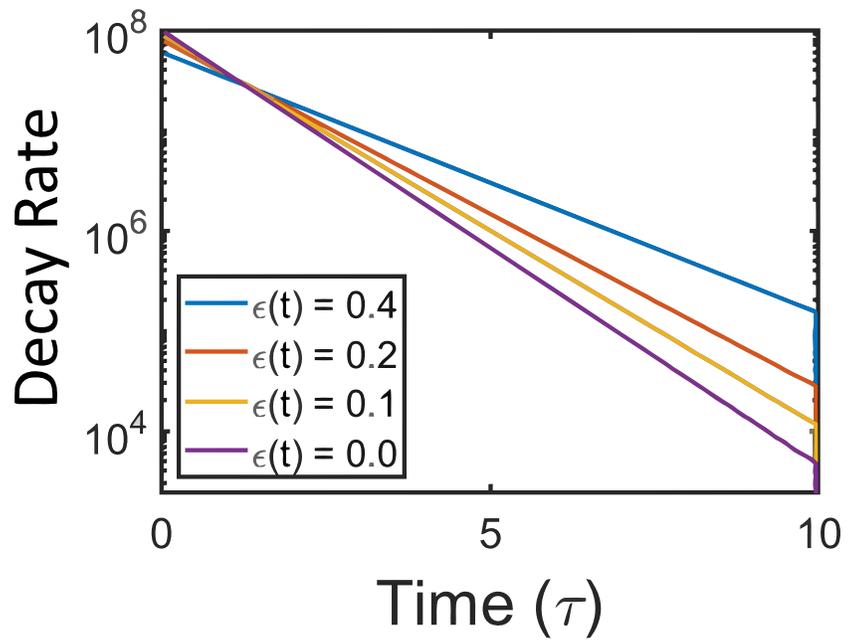} 
\caption{Simulation of the decay rate of $10^8$ excited particles as a function of time normalized to the lifetime of the particle in the indide view for different constant values of the probability $\epsilon(t)$  that the particle will remain in the excited state at each decay event that occurs in the outside view.} 
\label{t_0}
\end{figure}

Therefore, it is important to determine whether or not a constant value of $\epsilon$ is consistent with quantum kladodynamics before considering its implications for Everett's formulation of EQM and extensions thereof. We carried out such a simulation with Model 0 defined in Table \ref{Models}.  The results are shown in Fig. \ref{t_0}, which compares the logarithm of the results of simulations of the decay rate as a function of time for different values of $\epsilon$ with that of Model 0 in Table \ref{table2} ($\epsilon = 0$). 

The fact that the logarithms of the decay rates are linear on a log scale shows that they are consistent with Eq. \ref{Delta N(t)}.  The slopes of these lines agree quantitatively with Eq. \ref{AB}, which shows that reality splitting is consistent with decay of the lowest excited state to the ground state provided that $\epsilon = 1 - \tau_A/\tau_B = 1 - \Gamma_B/\Gamma_A$. 

In 1996 Oates et al.\cite{Oates} presented a graph that compared the results of two lower accuracy and two higher accuracy measurements of the lifetime of the 3p$^2$P$_{3/2}$ state of neutral sodium with the results of three high accuracy calculations.  According to the graph, the largest value from a high accuracy measurement was $\tau_A = 16.25$ ns, which was reported in 1995 by Volz et al. \cite{Volz03}. Also according to the graph, the lowest theoretical value was $\tau_B = 16.19$ ns, which was reported 1994 by T. Brage et al. \cite{Brage}.

From the point of view of CQM, this result is a conservative confirmation of a portion of CQM at the 0.4\% level.  From the point of view of reality splitting, this result places a conservative upper limit on $\epsilon$ for the 3p$^2$P$_{3/2}$ state of neutral sodium.   Specifically,
\begin{equation} \label{NewPhysics}
\epsilon_{3p^2P_{3/2}} \le 1 - \tau_{A}/\tau_{B} \lessapprox 0.004.  
\end{equation}

This means that if we have a single observers on a single branch at the start of a measurement, reality branching requires 240 observers who record a count during the measurement period for every observer who does not record a count during the same measurement period.  Whether each observer is on a separate branch or all observers recording the same result are grouped on a single branch is not important at this point.  What is important is that fractions of an observer are not consistent with the Standard Model even though superpositions of a single observer in multiple states are consistent with that model. 

There are two ways to extend EQM based on the above result.  First we can continue with the DeWitt picture but replace a monotonically deceasing $\epsilon(t)$ with a constant $\epsilon$. This picture of reality branching is consistent with Eq. \ref{DeWitt4}, but it violates Everett's definition of Process 2 as, \lq\lq The continuous, deterministic change of state of an isolated system with time according to a wave equation $\partial \psi/\partial t = A \psi$, where $A$ is a linear operator" because $m(a(t)) = a^*(t)a(t) = \epsilon > 0$.  Therefore, 
\begin{equation}
||a(t)|| = \sqrt{\epsilon},
\end{equation}
which is also a constant for times $t$ that satisfy $t_1 < \min(t_d, t_2)$ in Eq. \ref{DeWitt4}.  However, prior to time $t = t_1$, 
\begin{equation}
||a(t)|| = 1.
\end{equation}
This introduces a discontinuous change of state at time $t = t_1$. 

Furthermore, it introduces a branch discontinuity at time $t = t_d$ if a decay event is recorded at time $t_d$.  In other words this approach, which is equivalent to treating $\epsilon$ as a new constant of nature does not solve the problems that Everett was trying to solve with his reformulation of CQM.  Instead, it replaces a discontinuous change in the coefficients of the eigenfunctions in a superposition with a discontinuous change from the number of baryons in the initial state to the number of baryons in the final state. 

In an alternative approach to extending EQM to excited state decay, $\tau_A$ is a new constant of nature for each excited state of a quantum system relative to a fundamental force in the Standard Model and an excited-state decay is treated as an instantaneous perturbation at time $t_d$ .  In this case $||a_0(t)|| = 0$ and $||a_0(t)|| = 1$ for $t_1 < t_d$ and $||a_0(t)|| = 1$ and $||a_0(t)|| = 0$ for $t > t_d$. 

This approach moves the discontinuity at $t_1$ to $t_d$ but retains the discontinuous change in the baryon number.  In either case, an infinitely rapid \lq\lq collapse of wave functions" \cite{Everett02} has been replaced by an infinitely rapid \lq\lq fracture of branches" without any description of the laws of quantum kladodynamics that must exist to govern branching. Furthermore, the cost of this trade off is not negligible since it violates cherished conservation laws.  In light of this trade off, EQM is implausible unless it is inevitable.  

\section*{Discussion}

Everett's acceptance of this tradeoff suggests that he believed that it was necessary to eliminate the wavefunction discontinuity built into the \lq\lq external observer" formulation of quantum mechanics, based on Process 1". If so this may arise from faulty logic implicit in the formulation, which is based on two independent assumptions and a definition, which is really just a third assumption. 

\paragraph{The first assumption}\hspace{-0.4 cm} is \lq\lq that a wave function that obeys a linear wave equation everywhere and at all times supplies \textbf{a complete mathematical model for every isolated physical system without exception}."  This is an assumption that may or may not be correct.  It seems quite plausible for small microscopic systems.  But the complexity of some larger systems (up to and including reality) might preclude a description in terms of a linear wavefunction in the same way that the complexity of the general polynomial equation of degree five or higher with arbitrary coefficients precludes a solution in terms of radicals.

\paragraph{The second assumption}\hspace{-0.4 cm} is that \lq\lq all elements of the super- position exist simultaneously, and \textbf{the entire process is quite continuous}." This is just an assumption, but combined with the first assumption, it strongly constrains reality since it implicitly assumes that time is a mathematical continuum.   

If nature does not preclude discontinuities, there is certainly nothing wrong with including them in formulations of quantum mechanics. But even if nature does preclude discontinuities, there is nothing wrong with including them in the \lq\lq external observer" formulation of CQM because it is just a mathematical idealization based on continuum time.  

By supporting discontinuities, the \lq\lq external observer" formulation is powerful enough to describe abrupt transitions on all pertinent time scales, but the fact that it supports discontinuities does not mean that they exist in real phenomena.  If nature does preclude discontinuities, it is still the case that transition times between the initial and final states of many quantum systems are so abrupt that they appear to be discontinuous. For instance, the lifetime of a W$^-$ boson is of the order of $t_W = 10^{-25}$ s during the transition of a neutron into a proton, electron, and anti-neutrino.  

When transition dynamics are not well approximated as abrupt, it is possible to break the transition down into more abrupt steps as in quantum chemistry until the results become insensitive to step size. The external-observer formulation of quantum mechanics is not the problem.  Like so many other aspects of quantum mechanics. It's an approximation, not a complete description of reality with a wavefunction.   

\paragraph{The third assumption}\hspace{-0.4 cm} is hidden in Everett's definition of a good measurement.  He implicitly assumes that \textbf{the only way to avoid the \lq\lq discontinuous change} brought about by the observation of a quantity with eigenstates $\phi_1, \phi_2, \dots$, in which the state $\psi$ will be changed to the state $\phi_j$ with probability $|(\psi,\phi_j)|^2$" is to split reality into independent branches on each of which an observer who was aware of the existence of the superposition described by the state $\psi$ prior to the observation is aware of a different eigenstate of the superposition following each observation. 

Reality branching in EQM is built on these three assumptions.  But they are just assumptions.  Reality branching as a concept is an inevitable consequence of these three assumptions if these three assumptions are applied to reality instead of to a mathematical construct. However, since they are assumptions about reality, they do not prove anything about reality, much less the reality of the branching process. 

But the reality of the situation is even worse. The third assumption is patently false. A much more straight forward way to eliminate discontinuous changes wherever they appear is to understand that Process 1, which accommodates discontinuous changes, is a mathematical model of abrupt changes, and neither implies or reject the reality of discontinuous changes in nature. This is what I thought I learned when I studied quantum mechanics in 1968.  

In fact, our results show \lq\lq that a wave function that obeys a linear wave equation everywhere and at all times supplies" and that splits reality into independent branches as EQM describes must support splits that appear just as discontinuous or just as continuous as the collapse that he claims to eliminate. 

As already mentioned, what Everett did was to replace the discontinuous or abrupt collapse of the wave packet by the discontinuous or abrupt split of the branch.  Maybe this is why he \lq\lq never clearly spelled out how his theory was supposed to work"\cite{Schlosshauer} by writing the equations that described the actual splitting event.        

Everett also incorporated the observer (measurement system) into his formulation of quantum mechanics and derived a number of information theoretic results from this formulation. An interesting test of this aspect of EQM is the extent to which quantum physicists and chemists are actually using observer states like $\psi^{S+0'}[... \alpha_i]$ in designing and interpreting the results of measurements, as opposed to using them to support or refute the reality of quantum kladodynamics. Reality branching is irrelevant to answering this question.   

\section*{Conclusion}
 
\begin{enumerate}
  \item Everett's formulation as stated by him is not general enough to describe the spontaneous decay of an excited particle, but this is not a fatal flaw. 
  \item No DeWitt-like extensions of EQM are consistent with the experimental facts of tritium decay.
  \item If quantum kladodynamics is a valid formulation of reality (that is, if reality branching is a property of the universe), then it is a small effect for electronic transitions in sodium and other simple atoms for which accurate theoretical calculations and experimental measurements have been compared. 
  \item The statement that reality branching is an inevitable consequence of quantum mechanics is false.  In fact, it is not even a plausible consequence of quantum mechanics but only an inevitable consequence of a false premise. 
\end{enumerate}

\end{document}